# B PHYSICS AT LHCb

Monica Pepe Altarelli and Frederic Teubert

*PH Department, CERN, Geneva, Switzerland*
*E-mail: Monica.Pepe.Altarelli@CERN.CH*
*Frederic.Teubert@CERN.CH*

LHCb is a dedicated detector for *b* physics at the LHC. In this article we present a concise review of the detector design and performance together with the main physics goals and their relevance for a precise test of the Standard Model and search of New Physics beyond it.

1. **Introduction**

LHCb is a dedicated *b* and *c*-physics precision experiment at the LHC that will search for New Physics (NP) beyond the Standard Model (SM) through the study of very rare decays of charm and beauty-flavoured hadrons and precision measurements of CP-violating observables. At present, one of the most mysterious facts in particle physics is that, on the one hand, NP is expected in the TeV energy range to solve the hierarchy problem, but, on the other hand, no signal of NP has been detected through precision tests of the electroweak theory at LEP, SLC, Tevatron or through flavour-changing and/or CP-violating processes in *K* and *B* decays. In the last decade, experiments at *B* factories have confirmed the validity of the SM within the accuracy of the measurements. The domain of precision experiments in flavour physics has been extended from the rather limited kaon sector to the richer and better computable realm of *B* decays. The main conclusion of the first generation of *B*-decay experiments can be expressed by saying that the Cabibbo-Kobayashi-Maskawa (CKM) description (1)(2) of flavour-changing processes has been confirmed in $b \rightarrow d$ transitions at the level of 10-20% accuracy (3). However, NP effects can still be large in $b \rightarrow s$ transitions, modifying the $B_s$ mixing phase $\phi_s$ measured from $B_s \rightarrow J/\psi(\mu\mu)\phi$ decays (4), or in channels dominated by other loop diagrams, like, for example, the very rare decay $B_s \rightarrow \mu^+\mu^-$, e.g. via Higgs penguin diagrams (5)(6). Therefore, the challenge of the future $b$ experiments is to widen the range of measured decays, reaching channels that are strongly suppressed in the SM and, more generally, to improve the precision of the measurements to achieve the necessary sensitivity to NP effects in loops. LHCb will extend the $b$-physics results from the $B$ factories by studying decays of heavier $b$ hadrons, such as $B_s$ or $B_c$, which will be copiously produced at the LHC. It will complement the direct search of NP at the LHC by providing important information on the NP flavour structure through a dedicated detector, optimized for this kind of physics.

The main physics goals, which will be addressed in this review, include:

- Precision measurements of the CKM matrix elements including a search for possible inconsistencies in measurements of the angles and sides of the unitarity triangles using suitable decays, which proceed through different types of diagrams. A comparison of results from decays dominated by tree-level diagrams with those that start at loop level can probe the validity of the SM.
- Measurements of processes that are strongly suppressed in the SM and are poorly constrained by existing data, particularly in $b \rightarrow s$ transitions. Such processes, which could be enhanced through the impact of NP, include measurements of the $B_s$ mixing phase $\phi_s$ and of the very rare decay $B_s \rightarrow \mu^+\mu^-$.

## 2. *b* Physics at the LHC: Environment, Background, General Trigger Issues

The LHC will be the world's most intense source of $b$ hadrons. In proton-proton collisions at $\sqrt{s} = 14$ TeV, the $b\bar{b}$ cross section is expected to be ~500 $\mu$b producing $10^{12}$ $b\bar{b}$ pairs in a standard ($10^7$s) year of running at the LHCb operational luminosity of $2 \times 10^{32} \text{cm}^{-2} \text{sec}^{-1}$. The large centre of mass energy means that a complete spectrum of *b* hadrons will be available, including $B_{(s)}, B_{(c)}^+$ mesons and $\Lambda_b$ baryons. However, less than 1% of all inelastic events contain *b* quarks, hence triggering is a critical issue.

At the nominal LHC design luminosity of $10^{34} \text{cm}^{-2} \text{sec}^{-1}$, multiple *pp* collisions within the same bunch crossing (so-called pile-up) would significantly complicate the *b* production and decay-vertex reconstruction. For this reason the luminosity at LHCb will be locally controlled by appropriately focusing the beam to yield a mean value within $L = 2-5 \times 10^{32} \text{cm}^{-2} \text{sec}^{-1}$, at which most events have a single *pp* interaction. This matches well with the expected LHC conditions during the start-up phase. Furthermore, running at relatively low luminosity reduces the detector occupancy of the tracking systems and limits radiation damage effects.

The dominant $b\bar{b}$ production mechanism at the LHC is through gluon fusion in which the momenta of the incoming partons are strongly asymmetric in the laboratory frame. As a consequence, the centre of mass energy of the produced $b\bar{b}$ pair is boosted along the direction of the higher momentum gluon, and both *b* hadrons are produced in the same forward (or backward) direction. The detector is therefore designed as a single arm forward spectrometer covering the pseudorapidity range $1.9 < \eta < 4.9$, which ensures a high geometric efficiency for detecting all the decay particles from one *b* hadron together with a decay particle from the accompanying *b* hadron to be used as a flavour tag. A modification to the LHC optics, displacing the interaction point by 11.25 m from the centre, has permitted maximum use to be made of the existing cavern by freeing 19.7 m for the LHCb detector components.

A detector design based on a forward spectrometer offers further advantages: *b* hadrons are expected to have a hard momentum spectrum in the forward region; their average momentum is ~ 80 GeV/*c*, corresponding to approximately 7 mm mean decay distance, which facilitates the separation between primary and decay vertices. This property, coupled to the excellent vertex resolution capabilities, allows proper time to be measured with a few percent uncertainty, which is crucial for studying CP violation and oscillations with $B_s$ mesons, because of their high oscillation frequency. Furthermore, the forward, open geometry allows the vertex detector to be positioned very close to the beams and facilitates detector installation and maintenance. In particular, the silicon detector sensors, housed, like Roman pots, in a secondary vacuum, are positioned within ~8 mm from the beam during stable running conditions. They are split in two halves that are retracted by ~3 cm during injection.

Figure 1 illustrates the LHCb acceptance in the plane ($\eta$, $p_T$) of the *b* hadrons in comparison to that of ATLAS and CMS: ATLAS and CMS cover a pseudorapidity range of $|\eta| < 2.5$ and rely on high-$p_T$ lepton triggers. LHCb relies on much lower $p_T$ triggers, which are efficient also for purely hadronic decays. Most of the ATLAS and CMS *b*-physics programme will be pursued during the first few years of operation, for luminosities of order $10^{33}$cm$^{-2}$sec$^{-1}$. Once LHC reaches its design luminosity, *b* physics will become exceedingly difficult for ATLAS and CMS due to the large pile-up (~20 interactions per bunch crossing, on average), except for very few specific channels characterized by a simple signature, like $B_s \rightarrow \mu\mu$.

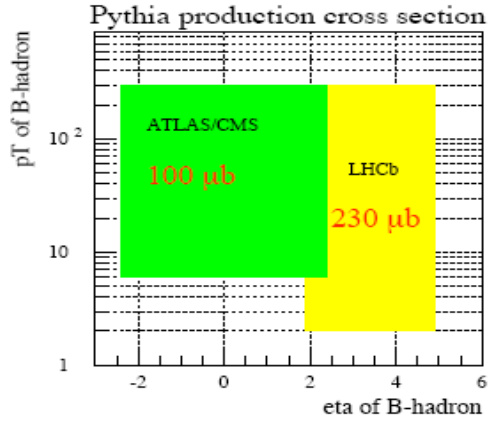

Fig. 1. *b* hadron transverse momentum $p_T$ as a function of the pseudorapidity $\eta$, showing the ($\eta$, $p_T$) regions covered by ATLAS and CMS, compared to that covered by LHCb

## 3. Detector Description and Performance

The key features of the LHCb detector are:
• A versatile trigger scheme efficient for both leptonic and hadronic final states, which is able to cope with a variety of modes with small branching fractions;
• Excellent vertex and proper time resolution;
• Precise particle identification (ID), specifically: hadron $\pi/K$ separation, and lepton $e/\mu$ ID;
• Precise invariant mass reconstruction to efficiently reject background due to random combinations of tracks. This implies a high momentum resolution.

A schematic layout is shown in Fig. 2. It consists of a vertex locator (VELO) (7)(8), a charge particle tracking system with a large aperture dipole magnet (9), aerogel and gas Ring Imaging Cherenkov counters (RICH) (10)(8), electromagnetic (ECAL) and hadronic (HCAL) calorimeters (11) and a muon system (12).

In the following, the most salient features of the LHCb detector are described in more detail.

### 3.1. Trigger

One of the most critical elements of LHCb is the trigger system (13). At the chosen LHCb nominal luminosity, taking into account the LHC bunch crossing structure, the rate of events with at least two particles in the LHCb acceptance is ~10 MHz (instead of the nominal 40 MHz LHC crossing rate). The rate of events containing $b$ quarks is ~100 kHz while the rate of events containing $c$ quarks is much larger (~600 kHz). However the rate of 'interesting' events is just a very small fraction of the total rate (~ Hz), hence the need for a highly selective and efficient trigger.

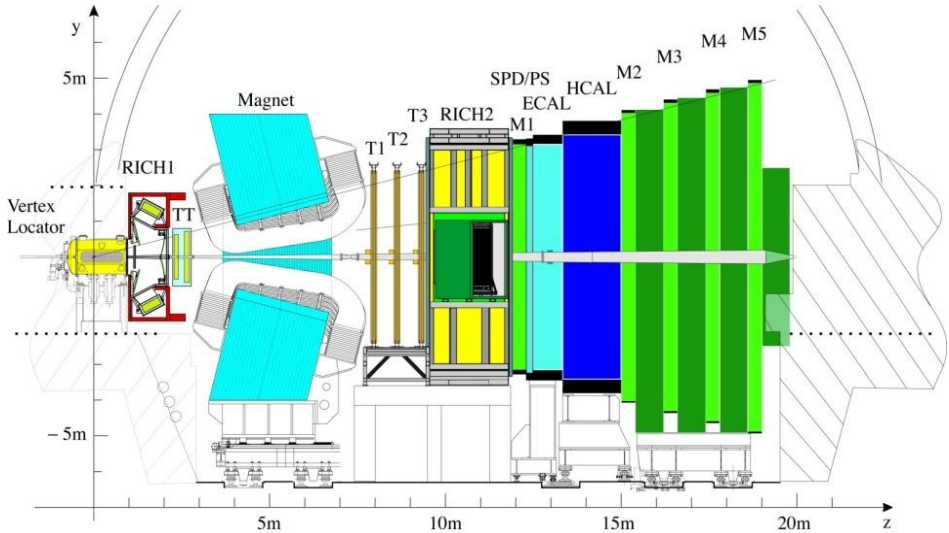

Fig. 2. Side view of the LHCb detector showing the Vertex Locator (VELO), the dipole magnet, the two RICH detectors, the four tracking stations TT and T1-T3, the Scintillating Pad Detector (SPD), Preshower (PS), Electromagnetic (ECAL) and Hadronic (HCAL) calorimeters, and the five muon stations M1-M5.

The LHCb trigger exploits the fact that *b* hadrons are long-lived, resulting in well separated primary and secondary vertices, and have a relatively large mass, resulting in decay products with large $p_T$.

The LHCb trigger consists of two levels: Level0 (L0) and High Level Trigger (HLT). L0, implemented on custom boards, is designed to reduce the input rate to 1 MHz at a fixed latency of 4 µs. At this rate, events are sent to a computer farm with up to ~2000 multi-processor boxes where several HLT software algorithms are executed. The HLT, which has access to the full detector information, reduces the rate from 1 MHz to ~2kHz.

L0, based on calorimeter and muon chamber information, selects muons, electrons, photons or hadrons above a given $p_T$ or $E_T$ threshold, typically in the range 1 to 4 GeV. A pile-up trigger system is also foreseen: two dedicated silicon disks located upstream of the VELO are used to reconstruct the longitudinal position of the interaction vertices and reject events with two or more such vertices, thus reducing the processing time on the remaining events as well as simplifying the offline analysis. However, if two muon candidates are found with $\sum p_T$>1.3 GeV/*c,* the pile-up is overridden; for this reason the L0 efficiency for *B* decays with two muons is very high independently of the instantaneous luminosity. The L0 hadron trigger occupies most of the bandwidth (~700 kHz) and is unique within the LHC experiments. The muon triggers (single and double) select ~200 kHz, while the rest of the bandwidth is due to the electromagnetic calorimeter triggers. Typically, the L0 efficiency is ~50% for hadronic channels, ~90% for muon channels and ~70% for radiative channels, normalized to offline selected events.

The HLT algorithms are designed to be fast and simple to understand in terms of systematic uncertainties. This is realized by reconstructing for each trigger only a few tracks, which are used for the final decision. The HLT comprises several paths (alleys) to confirm and progressively refine the L0 decision, followed by inclusive and exclusive selections. The choice of the alley depends on the L0 decision. For instance, if L0 triggers because of a hadron candidate (L0-*h*), a unique feature of LHCb, the trigger rate is further reduced by confirming the L0-*h* using the VELO. The candidate is then matched to a segment in the tracking stations and if it has $p_T$ and impact parameter above given thresholds, the event is sent to the inclusive/exclusive HLT selections. If these conditions are not satisfied, but a second 'companion' track is found in the VELO, which forms a good secondary vertex with the previous candidate, the event is also sent to the inclusive/exclusive HLT selections. Similar algorithms are applied to the other L0 decisions: L0-*μ*, L0-*e* and L0-*γ*. The average execution time is few ms, which matches with the expected size of the CPU farm. The total trigger rate after the HLT is ~2 kHz, a relatively high rate that also includes calibration samples to be used to understand the detector performance. The total rate can be approximately subdivided in ~200 Hz from the exclusive selections, ~300 Hz from the inclusive *D\** selection, also usable for particle identification efficiency studies, ~600 Hz from the inclusive di-muon selection without impact parameter cuts, also usable to calibrate the proper time distribution, and ~900 Hz from the inclusive muon selection for calibration of the trigger and tagging performance.

Overall, the combined trigger efficiency (L0+HLT) is expected to be ~30% for hadronic channels ($B_{(s)} \rightarrow hh$, $B_s \rightarrow D_s h$, etc), ~40% for channels with photons ($B \rightarrow K^*\gamma$, $B_s \rightarrow \phi\gamma$, etc…), ~70% for channels with muons ($B_s \rightarrow J/\psi(\mu\mu)\phi$, etc…), and ~90% for simple channels as $B_s \rightarrow \mu\mu$, normalized to offline selected events.

## *3.2. VELO and Tracking System*

The LHCb tracking system consists of a warm dipole magnet, which generates a magnetic field integral of ~4 Tm, four tracking stations (14)(15) and the VELO. The first tracking station located upstream of the magnet consists of four layers of silicon strip detectors. The remaining three stations downstream of the magnet are each constructed from four double-layers of straw tubes in the outer region, covering most (~98%) of the tracker area, and silicon strips in the area closer to the beam pipe (~2%). However, ~20% of the charged particles traversing the detector go through the silicon inner tracker, due to the forward-peaked multiplicity distribution. The expected momentum resolution, displayed in Fig.3a as a function of momentum, increases from $\delta p/p \sim 0.35\%$ for low momentum tracks to 0.55% at the upper end of the momentum spectrum.

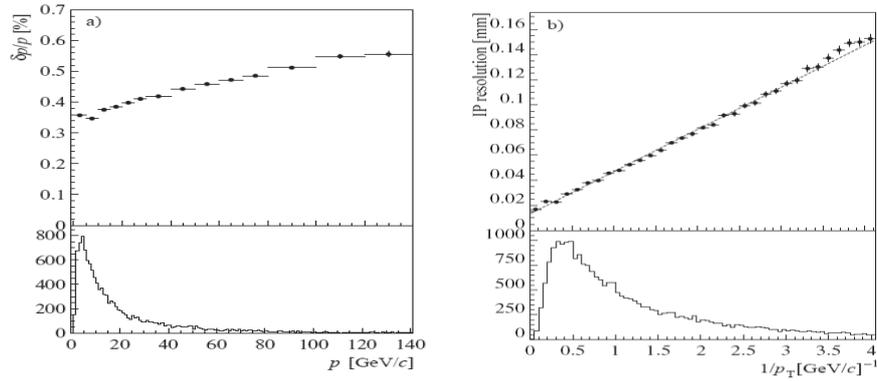

Fig.3. Resolution of the reconstructed track parameters at the track production vertex : a) momentum resolution as a function of track momentum, b) impact parameter resolution as a function of $1/p_T$. For comparison, the $p$ and $1/p_T$ spectra of *b*-decay particles are also shown in the bottom plots.

This translates into an invariant mass resolution of $\delta M \sim 20$ MeV/$c^2$ for $B_{(s)}$ decays into two charged tracks, like $B_s \to \mu\mu$, substantially better than in the General-Purpose detectors at LHC.

The VELO consists of 21 stations, each made of two silicon half disks, which measure the radial and azimuthal coordinates. The VELO has the unique feature of being located at a very close distance from the beam line (~0.8 cm), inside a vacuum vessel, separated from the beam vacuum by a thin aluminum foil. This allows an impressive vertex resolution to be achieved, translating, for instance, in a proper time resolution of ~36 fs for the decay $B_s \to J/\psi\,(\mu\mu)\phi$, i.e. a factor of ten smaller than the $B_s$ oscillation period and a factor of two better than in the General-Purpose detectors. The resolution on the impact parameter can be parameterized as $\delta IP \sim 14\mu m + 35\mu m/p_T$, as shown in Fig.3b.

### 3.3. *Particle Identification*

Particle identification is provided by the two RICH detectors and the Calorimeter and Muon systems. The RICH system is one of the crucial components of the LHCb detector. The first RICH, located upstream of the magnet, employs two radiators, $C_4F_{10}$ gas and aerogel, ensuring a good $\pi/K$ separation in the momentum range from 2 to 60 GeV/$c$. A second RICH in front of the calorimeters, uses a $CF_4$ gas radiator and extends the momentum coverage up to ~100 GeV/$c$. The calorimeter system comprises a pre-shower detector consisting of 2.5 radiation length lead sheet sandwiched between two scintillator plates, a 25 radiation length lead-scintillator electromagnetic calorimeter of the shashlik type and a 5.6 interaction length iron-scintillator hadron calorimeter. The muon detector consists of five muon stations equipped with multiwire proportional chambers, with the exception of the centre of the first station, which uses triple-GEM detectors.

Electrons, photons and $\pi^0$s are identified using the Calorimeter system. The average electron identification efficiency in the ECAL acceptance extracted from $J/\psi \to e^+e^-$ decays is ~95% for a pion misidentification rate of ~0.7%. Muons are identified using the muon detector with an average efficiency in the acceptance extracted from $J/\psi \rightarrow \mu\mu$ decays of ~93% for a pion misidentification rate of ~1%.

The RICH system provides a clean separation of hadron types, as well as some separation between leptons and hadrons, which is used to improve the overall particle identification performance. An example of the ability to identify kaons using the RICH system is shown in Fig.4. By changing the cut on the likelihood difference between the "*K* hypothesis" and the "$\pi$ hypothesis", the pion misidentification rate can be reduced, thus improving the purity of the selected sample, at the cost of some loss in the kaon identification efficiency. The trade-off between efficiency and purity can be adjusted according to the needs of individual analyses. For instance, a strong reduction in pion contamination is crucial to be able to separate the decay $B_s \rightarrow D_s K$ from the 15 times more abundant $B_s \rightarrow D_s \pi$.

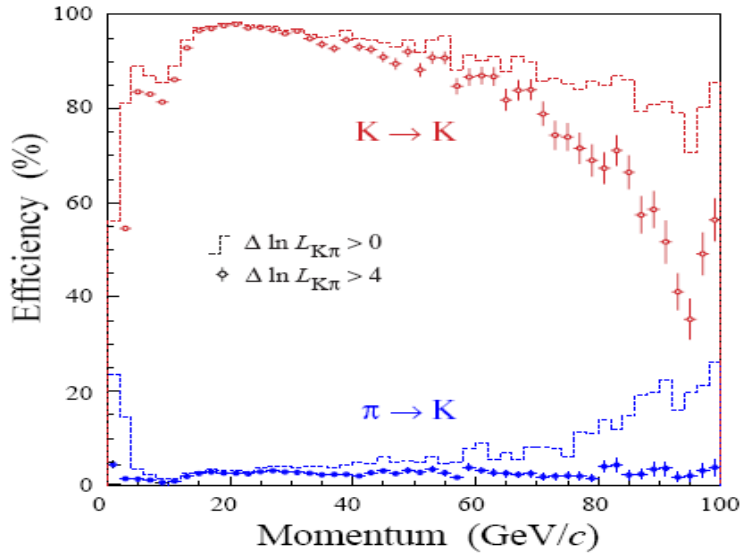

Fig.4. Kaon identification efficiency and pion misidentification rate as a function of momentum, for two different values of $\Delta \ln L_{K\pi}$, indicated by histograms and points, respectively.

## 4. Physics Objectives

The objective of the earliest running phase is to complete the commissioning of the sub-detectors and of the trigger, which includes time and space alignment, calibration of momentum, energy and particle identification. An integrated luminosity of ~0.5 fb$^{-1}$, which should be collected during the first year of physics running, will already allow LHCb to perform a number of very significant measurements, with the potential of revealing NP effects, such as the measurement of the $B_s$ mixing phase $\phi_s$, or the search of the decay $B_s \rightarrow \mu\mu$ beyond the limit set by CDF and D0. In subsequent years, the experiment will develop its full physics programme, and plans are to accumulate an integrated luminosity of ~10 fb$^{-1}$. Such a data sample will, for example, allow LHCb to improve the error on the CKM angle $\gamma$ by a factor of ~five, and probe NP in rare *B* meson decays with electroweak, radiative and hadronic penguin modes. In the following, after a brief introduction of the necessary formalism, some selected physics highlights are reviewed that are of particular relevance for NP discovery. A description of the other LHCb physics goals and sensitivity can be found in (8).

### 4.1. Introduction of Formalism

In the SM the unitary matrix $V_{CKM}$ arises from a misalignment of the flavour and mass eigenstate basis (1)(2): $D'=V_{CKM}D$, where

$$V_{CKM} = \begin{pmatrix} V_{ud} & V_{us} & V_{ub} \\ V_{cd} & V_{cs} & V_{cb} \\ V_{td} & V_{ts} & V_{tb} \end{pmatrix}$$

and $D'^T=(d',s',b')$ and $D^T=(d,s,b)$ are the flavour and mass eigenstates, respectively. After applying unitarity constraints and using the freedom of redefining the relative quark phases, the matrix $V_{CKM}$ depends on three mixing angles and one phase, which is the unique source of CP violation in the SM. The unitarity of the matrix implies $\sum_i V_{ij}V_{ik}^* = 0$ for $j \neq k$. Each of these six unitarity constraints can be seen as the sum of three complex numbers, closing a triangle in the complex plane. One of these relations, for *j=d* and *k=b* (*db* unitarity triangle) is of particular interest as it applies directly to *b* decays and is given by: $V_{ud}V^*_{ub}+V_{cd}V^*_{cb}+V_{td}V^*_{tb}= 0$. This relation defines the angles *α, β* and *γ* commonly used in the literature. Their general expression is:

$$\alpha = \arg\left(\frac{-V_{td}V_{tb}^*}{V_{ud}V_{ub}^*}\right), \quad \beta = \arg\left(\frac{-V_{cd}V_{cb}^*}{V_{td}V_{tb}^*}\right) \text{ and } \gamma = \arg\left(\frac{-V_{ud}V_{ub}^*}{V_{cd}V_{cb}^*}\right).$$

The unitarity relation with *j=u* and *k=t* (*ut* unitarity triangle) is of special relevance for the physics of the $B_s$ mesons, of particular interest for the LHCb experiment: $V_{ud}V^*_{td}+V_{us}V^*_{ts}+V_{ub}V^*_{tb} = 0$. It defines a triangle with angles very similar to those of the *db* unitarity triangle, except for a small shift: $\beta+\beta_s$ and $\gamma-\beta_s$, where $\beta_s$ is defined as:

$$\beta_s = \arg\left(\frac{-V_{cb}V_{cs}^*}{V_{tb}V_{ts}^*}\right).$$

A useful parameterization of the CKM matrix for phenomenological applications is that of Wolfenstein (16). This parameterization corresponds to an expansion in terms of $\sin\theta_c = \lambda \sim 0.22$, where $\theta_c$ is the Cabibbo angle, with three additional real parameters *ρ, η, A*. If one takes this expansion up to $O(\lambda^4)$, only three CKM elements have an imaginary part: $V_{td}$, $V_{ts}$ and $V_{ub}$ and in the SM all phases are proportional to *η*. One has:

$$\beta \approx -\arg(V_{td}) \approx \arctan\left(\frac{\eta(1-\lambda^2/2)}{1-\rho(1-\lambda^2/2)}\right)$$

$$\beta_s \approx \arg(V_{ts}) - \pi \approx \lambda^2\eta$$

$$\gamma \approx -\arg(V_{ub}) \approx \arctan(\frac{\eta}{\rho}).$$

In the SM the $B_{(s)} - \overline{B}_{(s)}$ transitions arise from box diagrams; the off-diagonal elements of the mass matrix $M_{12}=M^*_{21}$ (in the $B_{(s)}$, $\overline{B}_{(s)}$ basis) are given by the box diagram amplitude and are dominated by *t*-quark exchange. Additional contributions from NP are not excluded and could make the result deviate from the SM prediction. On the other hand, the off-diagonal element of the width matrix $\Gamma_{12}$ is given by the absorptive part of the box diagram amplitude and is dominated by real final states to which both $B_{(s)}$ and $\overline{B}_{(s)}$ can decay and is less sensitive to NP. In the SM and adopting the Wolfenstein parameterization, the phase of $M_{12}$ reduces to the box diagram phase $\phi_d = -2\arg\left(V_{td}^*V_{tb}\right) \approx 2\beta$ for the *B* system, and $\phi_s = 2\arg\left(V_{ts}V_{tb}^*\right) \approx -2\beta_s$ for the $B_s$ system. The SM values of $\phi_{d,s}$ are precisely predicted, therefore any significant deviation from the SM value would be a signal of NP. The decays of the neutral $B_{(s)}$ mesons to CP eigenstates like $B \to J/\psi K_s$ or $B_s \to J/\psi \phi$, which are dominantly produced by the $b \to c$ tree transition, are perfectly suited to measure the phase $\phi_{d,s}$ in the box diagram, as the only complex coupling involved is $V_{td(s)}$. Processes like $B \to D^{*\pm}\pi^{\mp}$ or $B_s \to D_s^{\pm}K^{\mp}$ measure the sum of the mixing phase $\phi_{d,s}$ and the $V_{ub}$ phase, i.e. the angle $\gamma$, through interference between the decays where the $B_{(s)}$ has or not oscillated.

These measurements will be illustrated in some more detail in the following sections.

## 4.2. Measurement of the $B_s$ Mixing Phase $\phi_s$

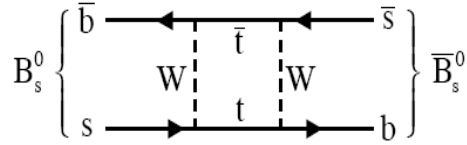

Fig.5. Box diagram contributing to the $B_s$ mixing in the SM.

The most promising channel (4) to measure the $B_s$ mixing phase $\phi_s$ is $B_s \to J/\psi(\to l^+l^-)\phi(\to K^+K^-)$. The final state $f = J/\psi\,\phi$ can be a CP eigenstate with eigenvalue $\eta_f = \pm 1$. The 'plus' sign applies to an orbital angular momentum $l=0$ or $2$, while the 'minus' sign is for $l=1$. What is to be measured is the time-dependent CP violating asymmetry defined as:

$$A_f^{CP}(t) = \frac{\Gamma_{\bar{B} \to f}(t) - \Gamma_{B \to f}(t)}{\Gamma_{\bar{B} \to f}(t) + \Gamma_{B \to f}(t)}.$$

A tagging method is needed to attribute each event to either $B_s$ or $\bar{B}_s$ decays. The time $t$ is the proper time of the decaying $B_s$ or $\bar{B}_s$ counted from the production time of the pair at $t=0$. The two interfering amplitudes are for a $B_s$ ($\bar{B}_s$) to decay directly into $f$ or to first mix into $\bar{B}_s$ ($B_s$) (see Fig.5) and then decay. The complex parameter that determines the interference pattern is $\lambda_f \equiv \frac{q}{p}\frac{\bar{A}_f}{A_f}$, where $q$ and $p$ are the complex numbers describing the state mixing, while $A_f$ and $\bar{A}_f$ are the decay amplitudes. For this particular process, $|q/p|=1$ (i.e. no CP violation in the $B_s$ - $\bar{B}_s$ oscillations) and $|\bar{A}_f/A_f|=1$ (i.e. no CP violation in the decay amplitudes) with very good approximation, so that $\lambda_f$ is a pure phase. In the SM, the decay phase $\phi_d$ from the ratio $\bar{A}_f/A_f$ is negligible, because the dominant contribution is from a diagram that is real. The remaining phase from $q/p$ is determined from the box diagram and is given by $\lambda_f \approx \eta_f e^{-i\phi_s}$ (CP violation in the interplay between the mixing and decay amplitudes). In this case the time-dependent CP violation amplitude reduces to:

$$A_{CP}(t) = -\frac{\eta_f \sin\phi_s \sin(\Delta m_s t)}{\cosh(\Delta\Gamma_s t/2) - \eta_f \cos\phi_s \sinh(\Delta\Gamma_s t/2)},$$

where $\Delta m_s$ and $\Delta\Gamma_s$ are the difference between the mass and widths of the two $B_s$ mass eigenstates, following the sign conventions of (17).

Flavour tagging dilutes the measured CP asymmetry through a factor $D = (1-2\omega_{tag})$, where $\omega_{tag}$ is the probability of having the wrong identification ($\omega_{tag} = 1/2$ if there is no tag). Flavour tagging is performed reconstructing the charge of the $b$ hadron accompanying the $B$ meson under study (opposite side tag) from its decay products, i.e. leptons, kaons, as well as the charge of the inclusive secondary vertex. Moreover $B_s$ ($B$) mesons can be tagged by exploiting the correlation with charged kaons (pions) produced in the fragmentation decay chain (same side tag). These taggers are combined in a neural network. The effective tagging efficiency ($\varepsilon_{eff} = \varepsilon_{tag}D^2$), which is the figure of merit for the tagging power, varies between 7-9% for $B_s$ ( 4-5% for $B$ ) (18).

The *J/ψ φ* final state is a sum of CP eigenstates and each contribution can be disentangled on a statistical basis. This is realized by performing an analysis based on the so-called transversity angle, defined as the angle between the positive lepton and the *φ* decay plane in the *J/ψ* rest frame (4). The phase $\phi_s$ is determined through a simultaneous fit to the proper time and transversity angle distributions, as well as to the proper time distribution of a control sample of $B_s \rightarrow D_s\pi$ decays, which is used to extract $\Delta m_s$ and the tagging efficiency. Recently, the CDF and D0 collaborations have reported a first indication of the mixing phase: $\phi_s \in [0.32, 2.82]$ at 68% C.L. and $\phi_s = 0.57^{+0.24}_{-0.30}$ using ~2k $B_s \rightarrow J/\psi \phi$ signal candidates at CDF and D0 (19) (20). The combined result (21) deviates from the SM prediction ($\phi_s$ =-0.04 radians) by ~3$\sigma$. LHCb expects (22) approximately 130k $B_s \rightarrow J/\psi\phi$ signal events in 2 fb$^{-1}$ of data with a background over signal ratio B/S~0.1. The phase $\phi_s$ can also be extracted from pure CP eigenstates, such as $B_s \rightarrow J/\psi\ \eta$ or $B_s \rightarrow \eta_c\phi$, which do not require an angular analysis, but for which the statistics is much lower. Combining all these modes, the expected statistical sensitivity on $\phi_s$ is 0.021 (0.009) radians with 2 (10) fb$^{-1}$ of data (22)(23), i.e. one (five) nominal year(s), which corresponds to a ~2$\sigma$ (~4$\sigma$) measurement assuming the SM value.

### *4.3. $B_s \rightarrow \phi\phi$ as a Probe for New Physics*

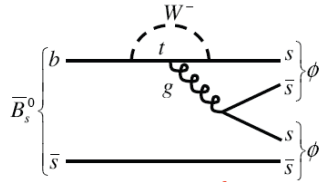

Fig.6. Penguin diagram contributing to the $B_s \rightarrow \phi\phi$ decay in the SM.

The discussion of the process $B_s \rightarrow \phi\phi$ proceeds along similar lines as for $B_s \rightarrow J/\psi\ \phi$. The final state *f* is again a sum of different CP eigenstates with $\eta_f = \pm 1$ and the relevant complex parameter is of the form $\lambda_f \equiv \frac{q}{p} \frac{\overline{A}_f}{A_f}$, where $p$ and $q$ are the same as before, while $A_f$ refers to the amplitude for $f = \phi\phi$. The crucial difference is however that in this case the dominant decay amplitude $A_f$ is a penguin diagram with top exchange (as shown in Fig.6), so that the decay phase $\phi_d$ is not negligible. Indeed it turns out that in the SM there is a complete cancellation between $\phi_s$ and $\phi_d$ (for the dominant penguin with top exchange) (24):

$$\phi_S^{SM}(B_S \to \phi\phi) \approx 2\arg(V_{ts}^* V_{tb}) - \arg(V_{tb}V_{ts}^* / V_{tb}^* V_{ts}) = 0.$$

Therefore this process is particularly sensitive to NP, which can introduce new CP-violating phases in the penguin decay and/or $B_s$ mixing. When combining with the decay $B_s \to J/\psi \phi$, one can disentangle NP contributions in $B_s$ mixing and decay. LHCb expects approximately 3.1k signal events in 2 fb$^{-1}$ of data with a background over signal ratio B/S<0.8 at 90% C.L. From the time-dependent angular distribution of flavour-tagged events, the phase $\phi_s$ can be measured with a statistical precision $\sigma(\phi_s) = 0.11$ (0.05) with 2(10) fb$^{-1}$ of data (25).

## 4.4. *Measurement of the Weak Decay-Phase $\gamma$ from Tree-Level Processes*

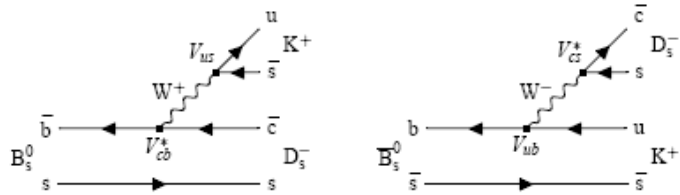

Fig.7. Feynman diagrams for $B_s \to D_s^- K^+$ and $\overline{B}_s \to D_s^+ K^-$ in the SM.

The weak phase $\gamma + \phi_s$ can be measured at LHCb from a time-dependent CP asymmetry analysis of the decay $B_s \to D_s^{\mp} K^{\pm}$ (see Fig.7) (26). As a $B_s$ and a $\overline{B}_s$ can both decay as $D_s^+ K^-$, there is interference between the $B_s$ decays where the $B_s$ has or not oscillated. The intrinsic theoretical uncertainty in the extraction of $\gamma$ is estimated to be ~0.1%, so that this is not a limiting factor for the measurement. Contrary to the equivalent channel in the $B$ system, i.e. $B \rightarrow D^{(*)+}\pi^-$, the decay amplitudes for the $b \rightarrow c$ and $b \rightarrow u$ transitions are both of $O(\lambda^3)$ and their ratio can be extracted from data, allowing a clean determination of the CP angle $\phi_s + \gamma$. The main issue is separating the decay $B_s \rightarrow D_s^{\mp} K^{\pm}$ from the very similar decay $B_s \rightarrow D_s^{\mp} \pi^{\pm}$, with a ~15 times larger branching ratio. Here, use of the RICH system crucially allows the background contamination to be reduced down to ~15%. The unique capability of the LHCb trigger to select hadronic decays is instrumental to collecting a sufficient statistics. It is estimated that LHCb will select 31k events for 10 fb$^{-1}$ with a combinatorial background to signal ratio B/S<0.2. The estimated precision on $\gamma$ for the same integrated luminosity is ~4.6° (i.e. a ~7% uncertainty), assuming that $\phi_s$ is known from the measurements described in the previous section (27) (28).

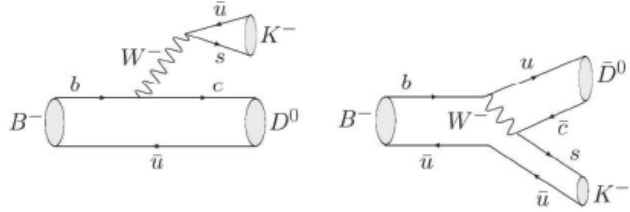

Fig.8. Feynman diagrams for $B^- \rightarrow DK^-$ and $B^- \rightarrow \overline{D}K^-$ in the SM. In the left diagram all colourless configurations of the $s$ and $\overline{u}$ are allowed (coloured-favoured diagram). In the right diagram the $W^-$ decay products are split to form the $\overline{D}$ and $K^-$ and therefore the colour is constrained by the colour of the initial $b$ and $\overline{u}$ quarks (colour-suppressed diagram).

Alternatively, $\gamma$ can be extracted from $B^{\pm}$ decays to open charm, as done by BaBar and Belle, where a precision of ~20° was achieved through this method (29) (30). From the experimental point of view, this analysis is much simpler as no flavour tagging or time-dependent analysis is required. Various methods using $B^{\pm} \rightarrow (D/\overline{D})K^{\pm}$ decays have been proposed. All of them are based on two key observations: 1. The decay $B^{\pm} \to (D/\overline{D})K^{\pm}$ can produce neutral $D$ mesons of both flavours via colour-favoured or colour-suppressed decays (see Fig.8) 2. Neutral $D$ and $\overline{D}$ mesons can decay to a common final state, for example through Cabibbo-favoured or doubly Cabibbo-suppressed Feynman diagrams (ADS method) (31) or through decays to CP eigenstates such as $K^+K^-$ or $\pi^+\pi^-$ (GLW method) (32) (33) (34). In the ADS case, the reversed suppression between $B$ and $D$ decays results in very similar amplitudes leading to a high sensitivity to $\gamma$. The relative phase between the two interfering amplitudes for $B^+ \to DK^+$ and $B^+ \to \overline{D}K^+$ is the sum of the strong and weak interaction phases, while in the case of $B^- \to DK^-$ and $B^- \to \overline{D}K^-$ the relative phase is the difference between the strong phase and $\gamma$. Therefore both phases can be extracted by measuring the two charge conjugate modes. The feasibility of this measurement crucially depends on the size of the ratio: $r_B=|A(B^+ \to DK^+)|/|A(B^+ \to \overline{D}K^+)|$. The $r_B$ value is given by the ratio of the CKM matrix elements involved $|V^*_{ub} V_{cs}|/|V^*_{cb} V_{us}|$ times a colour suppression factor, and is estimated to be in the range 0.1-0.2. The three-body final state of the $D$ meson can also be used through a Dalitz plot analysis that allows one to obtain all information required for the determination of $\gamma$ with a single decay mode (35) (36). In particular, LHCb has studied the $B^{\pm} \to D/\overline{D}(K\pi)K^{\pm}$ and $B^{\pm} \to D/\overline{D}(K\pi\pi)K^{\pm}$ modes. Assuming $r_B$ ~0.15, ~300k (2.5k) events are selected for 10 fb$^{-1}$ in the favoured (suppressed) modes, with B/S~0.3(~2). From these modes LHCb estimates a precision on $\gamma$ in the range (2-6)°, depending on the $D$ strong phase, with 10 fb$^{-1}$ of data (37) (38) (39).

A similar precision can be achieved with the neutral $B$-decay modes $B \to \overline{D}(K\pi)K^*(K\pi)$, $B \to D_{CP}(KK/\pi\pi)K^*(K\pi)$ and their charge conjugate modes, where $D_{CP}$ denotes the CP even eigenstate mode. These modes are self-tagged, as the $B$ flavour is determined by the $K$ charge in the $K^*$ decay, while the $D$ flavour is determined by the $K$ charge in the $D$ decay. It is expected (40) that the experiment will select ~19k events from the $B \to \overline{D}K^*$ mode summing over the four possible flavour combinations (with B/S~0.4-10) and ~3k events from the CP eigenstate channel (B/S<~4), assuming $r_B$~0.4, with 10 fb$^{-1}$ of data.

Combining all these different techniques, the final LHCb precision on $\gamma$ is expected to be ~2.4°, with 10 fb$^{-1}$ of data (41).

Figure 9 shows the direct measurement of $\gamma$ from the *B* factories ($\gamma = 83°\pm19°$) and the indirect determination (27) from the rest of the CP measurements available in 2006, ($\gamma = 64.1°\pm4.6°$), as well as a future scenario where the precision achieved by LHCb reveals NP effects as a disagreement between these two determinations.

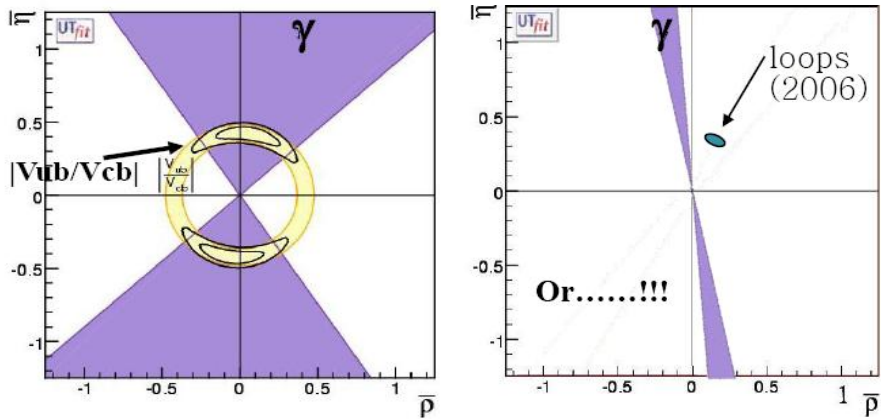

Fig. 9. a) Direct measurements of $\gamma$ using only "tree diagrams" as measured recently b) Comparison of a hypothetical LHCb direct measurement of $\gamma$ and a recent indirect determination. A discrepancy would indicate the presence of NP in loops affecting the indirect determination.

## 4.5. *Example of Radiative Penguins:* $B_s \rightarrow \phi\gamma$

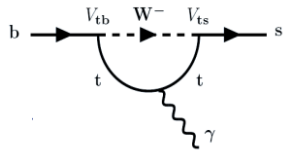

Fig.10. Example of Feynman diagram for $b \rightarrow s\gamma$.

The flavour-changing neutral current transition $b \rightarrow s\gamma$ (see Fig.10) is among the most valuable probes of NP models, in particular those where the flavour-violating chiral transition of the amplitude is not suppressed (42) (43) (44). Inclusive measurements of $B\to X_s\gamma$, which can be computed using perturbation theory, cannot be performed at LHCb. However, many exclusive measurements, like $B_s\to\phi(K^+K^-)\gamma$, are well suited to the LHCb detector capabilities. The CP asymmetry for this decay can be expressed as:

$$A_{CP}(t)=-\frac{C\cos(\Delta m_s t)+S\sin(\Delta m_s t)}{A^\Delta\sinh(\Delta\Gamma_s t/2)+\cosh(\Delta\Gamma_s t/2)},$$

The prediction for *S* is dominated by the electromagnetic dipole operator, and precisely known in the SM: $S_{\phi\gamma}$ = -0.1±0.1%. BaBar and Belle have measured a similar quantity in the $B\to K^*(K_s\pi^0)\gamma$ decay and it is consistent with the SM prediction (45) (46), however with large uncertainties: $\Delta S_{K^*\gamma}$~40% with samples of ~150 events. LHCb expects (47) to collect 36k (6k) events per fb$^{-1}$ in $B\to K^*(K^+\pi^-)\gamma$ ($B_s\to\phi(K^+K^-)\gamma$) decays with B/S~0.7 (0.9). In LHCb the $B\to K^*\gamma$ decays can be used as control channel, while the time-dependent CP asymmetry, which is sensitive to the photon polarization, can be measured in $B_s\to\phi\gamma$ decays, provided that the proper time resolution is sufficient to resolve the $B_s$ oscillations. The proper time resolution depends on the kinematics and topology of the particular $B_s$ candidates: mainly on the opening angle between kaons from the $\phi$ decay. LHCb should be able to measure $S_{\phi\gamma}$ with a precision of $\Delta S_{\phi\gamma}$ ~5% with 10 fb$^{-1}$ of data (48).

### 4.6. *Example of an Electroweak Penguin: $B\to K^*\mu\mu$*

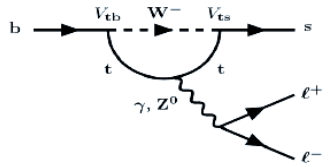

Fig. 11. Example of Feynman diagram of $b\to sl^+l^-$.

In the SM, the decays $b \to sl^+l^-$ cannot occur at tree level but only through electroweak penguin diagrams with small branching fractions, e.g. BR($B \to K^*\mu\mu$)~$1.2 \times 10^{-6}$ (see Fig.11). The $B \to K^*\mu\mu$ channel is well suited to searches for NP, because most NP scenarios make definite predictions for the forward-backward asymmetry $A_{FB}$ of the angular distribution of the $\mu$ relative to the $B$ direction in the di-muon rest frame as a function of the di-muon invariant mass $m_{\mu\mu}$. In particular, the value of $m_{\mu\mu}$ for which $A_{FB}$ becomes zero is predicted with small theoretical uncertainties and may thus provide a stringent test of the SM (49). The $B$ factory experiments BaBar and Belle have succeeded in measuring several observables for this exclusive decay: branching fraction, di-lepton angular asymmetry $A_{FB}$ vs di-lepton $q^2$, $K^*$ longitudinal polarization vs di-lepton $q^2$ and fits of the $d^2\Gamma/d\cos\theta\, dq^2$ distribution to extract the relevant Wilson coefficients, with a precision comparable to what LHCb can do with ~0.07 fb$^{-1}$ of data.

This exclusive decay can be triggered and reconstructed in LHCb with high efficiency (50), because of the clear di-muon signature and $K/\pi$ separation provided by the RICH detectors. Moreover, the invariant mass of the di-muon system is measured with an excellent resolution ($\sigma$~14 MeV/$c^2$). The selection criteria including the trigger have an efficiency of ~1%, leading to an expectation of ~7k signal events for an integrated luminosity of 2 fb$^{-1}$ and a background over signal ratio of ~0.5 in a ±50 MeV/$c^2$ mass window around the $B$ mass and ±100 MeV/$c^2$ window around the $K^*$ mass. About half of the background is estimated from an upper limit to the (not yet observed) non-resonant $B \to K^+\pi^-\mu^+\mu^-$ decay, which constitutes an irreducible background, while the other half originates from events with two semileptonic $B$ decays.

In addition to measuring $A_{FB}$ (51), LHCb will have enough statistics to extract the transversity amplitudes (52) (53) $A_0$, $A_{//}$ and $A_\perp$, through differential distribution defined in terms of $q^2$, the invariant mass square of the di-muon candidates, $\theta_l$, the angle between the $\mu^+$ and the $B$ in the di-muon rest frame, $\theta_K$, the angle between the $K$ and the $B$ in the $K\pi$ rest frame, and $\psi$, the angle between the normal to the planes defined by the di-lepton and $K\pi$ systems in the $B$ rest frame. Figure 12 shows an example of the LHCb sensitivity to the measurements of $A^{(2)}_T(q^2) = (|A_\perp|^2 - |A_{//}|^2)/(|A_\perp|^2 + |A_{//}|^2)$, which is sensitive to new sources of right-handed currents, and the fraction of $K^*$ polarization: $F_L(q^2) = |A_0|^2 / (|A_0|^2 + |A_\perp|^2 + (|A_\parallel|^2)$. Figure 12 also shows the SM NLO prediction with its uncertainty, and the predictions from the MSSM with Minimal Flavour Violation (MFV) and $\tan\beta = 5$. The fact that these measurements are sensitive to SUSY with low $\tan\beta$ values shows the complementarity with the $B_s \rightarrow \mu\mu$ measurement, which is sensitive to models with large $\tan\beta$, as explained in the next section. Unfortunately, in the more sensitive region $q^2 < 1$ $(\text{GeV}/c^2)^2$ the theory is not very reliable. In the theoretically favoured region, away from the photon pole, $q^2 > 1$ $(\text{GeV}/c^2)^2$, and below the charm resonances, $q^2 < 6(\text{GeV}/c^2)^2$, the resolution in $A^{(2)}_T$ is 0.42(0.16) and the resolution in $F_L$ is 0.016(0.007) with 2 fb$^{-1}$ (10 fb$^{-1}$) of integrated luminosity (54).

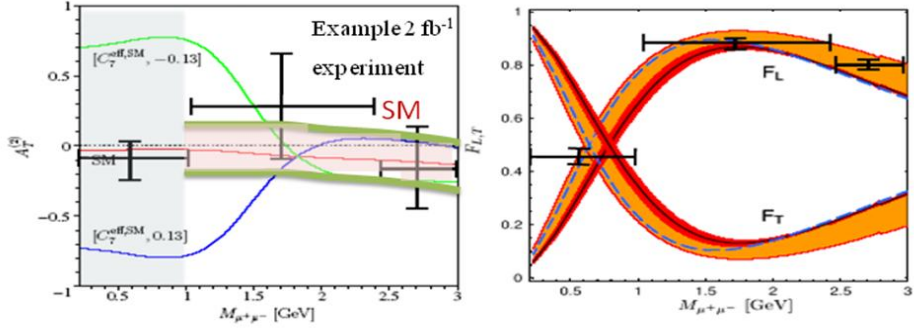

Fig. 12. $A^{(2)}_T$ (left) and $F_{L,T}$ (right) as a function of the di-muon mass. Also shown on the left plot is the SM NLO prediction for $A^{(2)}_T$ with its uncertainty (horizontal band), and the predictions from the MSSM with MFV and $\tan\beta = 5$ (green and blue lines). The grey-shaded region is excluded as theory is not reliable at low $q^2$. The right plot shows the SM prediction for $F_{L,T}$ and its uncertainty.

### 4.7. *Example of a Higgs-Penguin:* $B_s \rightarrow \mu^+\mu^-$

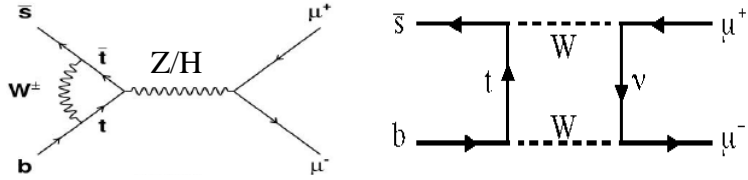

Fig.13. Examples of Feynman diagram for $B_s \rightarrow \mu^+\mu^-$.

The decay $B_s \rightarrow \mu^+\mu^-$ has been identified as a very interesting potential constraint on the parameter space of models for physics beyond the SM (5) (6). The upper limit to the $B_s \rightarrow \mu^+\mu^-$ branching ratio measured at the Tevatron (55) (56) is $4.7 \times 10^{-8}$ @90% C.L. The SM prediction is computed (57) to be BR($B_s \rightarrow \mu^+\mu^-$) = $(3.35 \pm 0.32) \, 10^{-9}$ using the latest measurement of the $B_s$ oscillation frequency at the Tevatron ($\Delta M_s = 17.8 \pm 0.1$ ps$^{-1}$) (58), which significantly reduces the uncertainties in the SM prediction. Within the SM, this decay is dominated by a "Z/Higgs-penguin" diagram (Fig.13 left), while the contribution from the "box" diagram (Fig.13 right) is suppressed by a factor $\sim (M_w/m_t)^2$. Moreover, the decay is helicity suppressed by a factor $\sim (m_l/m_{B_s})^2$, hence it is very sensitive to any NP with new scalar or pseudoscalar interactions, in particular to any model with an extended Higgs sector.

In the MSSM this branching ratio is known to increase as the sixth power of $\tan\beta = v_u/v_d$, the ratio of the two Higgs vacuum expectation values. Any improvement to this limit is therefore particularly important for models with large $\tan\beta$. For instance, Figure 14 shows the values of $\tan\beta$ and $M_A$, the mass of the CP odd neutral Higgs, preferred by a global fit to several measurements within one particular realization (59) of the MSSM. The best fit position is largely dominated by the present 3.4$\sigma$ discrepancy in the measured anomalous magnetic moment of the muon. As a consequence, if such a discrepancy is not due to a statistical fluctuation, a sizable enhancement of the branching ratio is expected in this model, i.e. BR($B_s \rightarrow \mu^+\mu^-$) $\sim 10^{-8}$.

The large background expected in the search for the decay $B_s \rightarrow \mu^+\mu^-$ is largely dominated by random combinations of two muons originating from two distinct $b$ decays. This background can be kept under control by exploiting the excellent tracking and vertexing capabilities of LHCb. Specific decays, such as $B_{(s)} \rightarrow h^+h^-$, where the hadrons ($h$) are misidentified as $\mu$, or $B_c \rightarrow J/\psi \, (\mu^+\mu^-) \, \mu\nu$, do not contribute to the background at a significant level, compared to the combinatorial background, due to the very low $\mu$ misidentification rate (~0.5%) and excellent invariant mass resolution ($\sigma \sim 20$ MeV/$c^2$). The LHCb trigger selects the signal very efficiently: the total trigger efficiency, L0+HLT, is larger than 90%, normalized to the offline selected events.

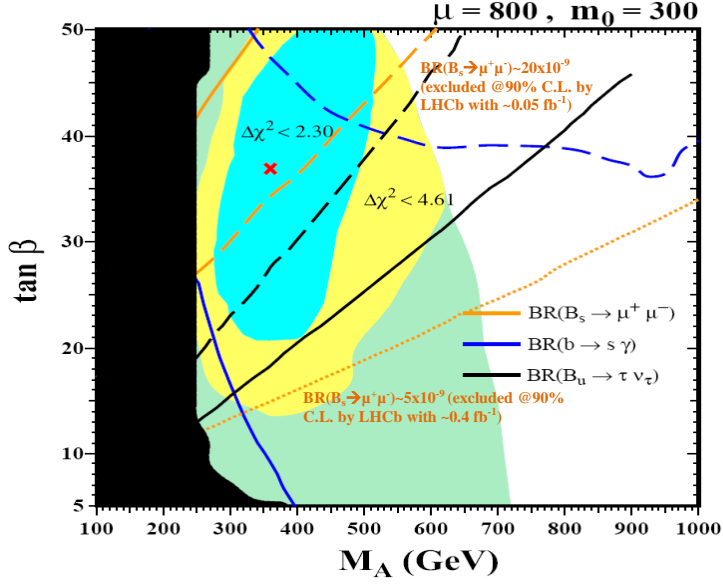

Fig. 14. Best fit and $\chi^2$ contours in the plane ($M_A$, tan$\beta$) from the fit in ref. (59) to several observables, including the anomalous magnetic moment of the muon. The orange lines indicate the excluded region when BR($B_s \to \mu^+\mu^-$)<$10^{-7}$ (2 $10^{-8}$) (5 $10^{-9}$).

The expected reach for exclusion at 90% C.L. of a given branching ratio if only background is observed, is also shown in Fig.14. LHCb has the potential to exclude the interesting region with very little luminosity (~0.4 fb$^{-1}$), and to make a 3$\sigma$ (5$\sigma$) observation (discovery) of the SM prediction with ~2 fb$^{-1}$ (~6 fb$^{-1}$) of data (60).

## 5. Conclusions and Outlook

The large $b\bar{b}$ production cross section at the LHC provides a unique opportunity to study in detail CP violation and rare *b* decays. In particular, production of $B_s$ mesons could play a crucial role in disentangling CP violation effects originating from NP. LHCb is an experiment characterized by a flexible and robust trigger system, sensitive to many *b* decays, including those with no lepton in the final state. It has a RICH system that allows a clean *π/K* separation over a wide momentum range, and a powerful vertex detector providing excellent decay-time resolution. LHCb can nicely complement the direct search of NP performed by ATLAS and CMS. If, as we all hope, NP phenomena will be observed directly by ATLAS and CMS, LHCb will become essential for elucidating the dynamics of such phenomena by looking for signals of virtual effects in flavour-changing and CP-violating processes.

## Acknowledgments

We would like thank our LHCb colleagues for providing the material discussed in this article, and, in particular, R.Forty, T.Nakada and O.Schneider for their valuable comments. We would also like to thank J.Ellis and K.A.Olive for having kindly provided Fig.14, J.Matias for discussions concerning the theoretical uncertainties displayed in Fig.12, and A.Lenz for his helpful comments on the manuscript.